\documentclass[a4paper]{article}
\usepackage{graphics}
\usepackage{amssymb}
\newenvironment{myfig2}[1]{\begin{figure}[#1] 
  \begin{center}
    \begin{tabular}{p{.465\textwidth}p{.01\textwidth}p{.465\textwidth}} }
    {\end{tabular}
  \end{center} 
 \end{figure}
}

\begin{document}
 
 
\begin{titlepage}
 
\hspace*{-5mm}Available on CMS information server
\hfill {\Large\bf CMS NOTE 1999/035}
 
\begin{center}
\includegraphics*{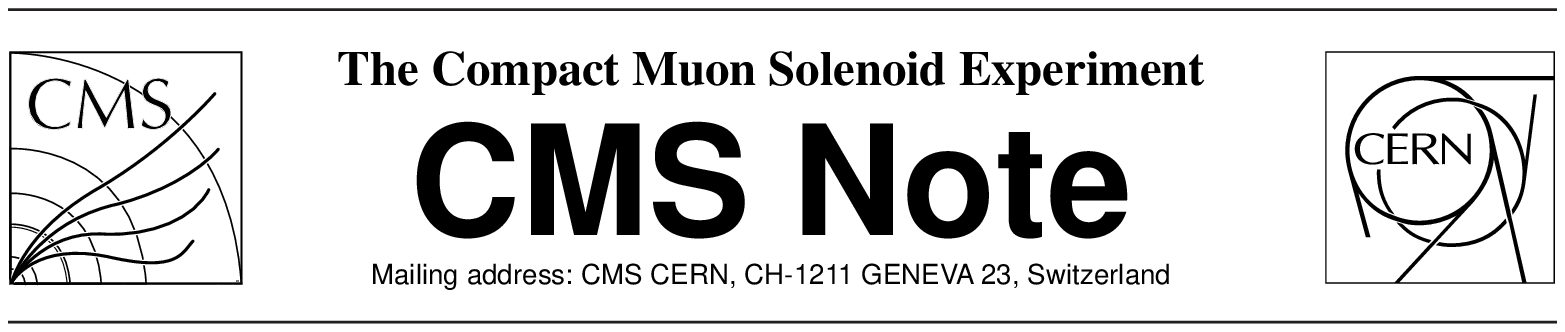} 
\end{center}
\begin{flushright}{\large\bf May 27, 1999}\end{flushright}
 
\vspace*{15mm}
 
\begin{center}
{\LARGE\bf On b- and $\tau$-multiplicities per event in SUSY (mSUGRA)
        and instrumental implications}
\end{center}
 
\vspace*{15mm}
 
  \begin{center}
    S.~Abdullin~$^{a)}$ \\
    IReS, Strasbourg, France.\\
    \vspace*{5mm}
     D.~Denegri~$^{b)}$ \\
     Centre d'Etudes Nucl\'{e}aire de Saclay, Gif-sur-Yvette,
      France
  \end{center}
 
\vspace{30mm}
  \begin{abstract}
We investigate the probability to find a b or  $\tau$
in SUSY production with the mSUGRA model. We find that in the entire 
parameter space the probability per event to find a b-jet of E$_T^b$ $>$ 50 GeV
 within CMS acceptance ($| \eta |$ $<$ 2.4) is significant for all tan$\beta$,
 varying from
a $\sim$ 10 \% level to 90 \%  depending on the m$_0$, m$_{1/2}$ region.
 The multiplicity of b-jets per event 
slightly increases with tan$\beta$. 
The probabilities per event to find a  $\tau$
with the same kinematical cuts 
is also significant, particularly in the region 
m$_{1/2}$ $>$ (1 -- 1.5)$\cdot$m$_0$, and it increases sharply with tan$\beta$.
These findings point to the central role  a microvertex
 device would play in case that SUSY (mSUGRA) is indeed realized in nature
and found at the LHC.
As the m$_0$, m$_{1/2}$ parameter space of largest $\tau$ and largest 
b-jet probability per event tend to be complementary, to have an efficient
high-performance three-layered microvertex device would be very sound and  
safe instrumental strategy in this context.
First investigations done in the context of the more general MSSM scenario
 confirm the findings based on mSUGRA.  
  \end{abstract}

\vspace*{20mm}
\hspace*{-8mm} \hrulefill \hspace{80mm} \hfill \\
$^{a)}$ On leave from ITEP, Moscow, Russia.
  Email: adullin@mail.cern.ch\\
$^{b)}$ Email: Daniel.Denegri@cern.ch

\end{titlepage}
\pagenumbering{arabic}
\setcounter{page}{1}

\section{Introduction}

One of the main purposes of the LHC collider is to search for the physics 
beyond the Standard Model (SM). One of the direction of this search is
to look for  superpartners of ordinary particles expected in 
Supersymmetric extensions of SM (SUSY). 
SUSY, if it exists, is expected to reveal itself at LHC firstly via an excess
 of (multilepton +) multijet + E$_{T}^{miss}$ final states
 compared to Standard Model (SM) expectations
\cite{reach_papers}. Spectacular and revealing structures are also possible
in $l^+l^-$ spectra indicative of SUSY production \cite{edge}.
It has been known for a long time that SUSY production will be accompanied
by an excess of b-jets, in part as $\tilde{t}_1$ and  $\tilde{b}_1$ are
expected to be the lightest among the squarks.
Recently \cite{large_tan} it has also been shown that with increasing
tan$\beta$ we should expect 
substantial increase of sparticle branching ratios to 
third generation particles due to increase of b and $\tau$ Yukawa couplings.
This implies a significant increase of the $\tau$ yield for large  tan$\beta$.
Both b's and $\tau$'s require a specific detection technique, relying mostly
on precision measurements of impact parameters. 

The main goal of this study is to
show the importance of good b-jet and tau detection in CMS
\cite{CMS}
for SUSY measurements.  
The importance of b-tagging in $h$ $\rightarrow$ $b\bar{b}$ detection is 
discussed in \cite{hbb,tracker_TDR}.

We are interested in evaluating the richness of b-production in SUSY,
whose detection relies on microvertexing, and of $\tau$ production
which requires calorimeter, tracker \cite{taudet} and microvertexing selection.
More emphasis is given to  $\tau$ production as less known and documented, 
especially at high tan$\beta$, 
but  b-production is more rewarding and instrumentally more demanding
 as it depends on a detection of several fairly soft ($\sim$ 1 -- 5 GeV/c) 
tracks with significant impact parameters in a jetty environment.

The paper is organized as follows.
We discuss the specific SUSY model employed in section 2.
 Phenomenological differences of low and high tan$\beta$ 
values are briefly discussed in section 3. Then in section 4 the difference 
between various domains of $m_{0}$, $m_{1/2}$ plane for the same tan$\beta$=35
is discussed.
The mSUGRA observations at high tan$\beta$ are somewhat generalized 
in case of the minimal extension of the SM (MSSM) in section 5. 
The main results of this study are summarized in section 6.

\section{Model employed}

The large number of SUSY parameters, even in the 
minimal extension of the SM (MSSM),  makes it difficult
to comprehend the variety of possible signals and signatures and
 to evaluate the
reach in various sparticle searches in general. So,
to have a better grasp of the situation,
we limit ourselves at present basically to the mSUGRA-MSSM model,
except for some specific cases explicitly mentioned.  This model
emanates from the MSSM, using Grand Unification Theory (GUT) assumptions
to limit the number of parameters 
(see more details in e.g. 
\cite{msugra}).
 
The mSUGRA model contains only five free parameters $:$
\begin{itemize}
 \item a common gaugino mass ($m_{1/2}$) ;
 \item a common scalar mass ($m_{0}$);
 \item a common trilinear interaction amongst the scalars ($A_{0}$);
 \item the ratio of the vacuum expectation values of the Higgs fields that
       couple to $T_{3} = 1/2 $  and $T_{3} = -1/2$ fermions ( tan$\beta$);
 \item a Higgsino mixing parameter $\mu$ which enters only through its
   sign ($sign (\mu)$).
\end{itemize}

 For a given choice of model parameters all
 the masses and couplings, and thus production cross sections
(up to structure function)
 and decay branching ratios are fixed. 

The mass of the lightest SUSY particle (LSP) which is 
$\tilde{\chi}_{1}^{0}$ in the R-parity conserving mSUGRA equals approximately
$\sim$ 1/2 of the $\tilde{\chi}_{2}^{0}$ mass. The mass of lightest chargino
 $\tilde{\chi}_{1}^{\pm}$ is almost
the same as that of the $\tilde{\chi}_{2}^{0}$. Isomass contours of 
$\tilde{\chi}_{1,2}^{0}$ and  $\tilde{\chi}_{1}^{\pm}$ and gluino behave
 gaugino-like, i.e. depend mainly on $m_{1/2}$. Masses of sleptons and squarks
depend on both $m_{0}$ and $m_{1/2}$.

Masses of squarks (especially of the first generation), gluinos, charginos and
neutralinos depend only weakly on tan$\beta$,
 A$_{0}$ or sign($\mu$) parameters. 
Masses of sleptons, stop and sbottom
have some dependence on these mSUGRA
parameters, in particular third generation sparticles on tan$\beta$;
masses of $\tilde{\tau}_1$,  $\tilde{t}_1$ and $\tilde{b}_1$ tend to be 
the lightest among the sleptons and squarks respectively with increasing 
 tan$\beta$.
Masses of Higgs bosons depend significantly on 
tan$\beta$, the mass
of the lightest scalar Higgs $h$ increases with tan$\beta$ 
and depends also on sign($\mu$), whilst 
masses of the heavy Higgses decrease dramatically with increasing tan$\beta$
in this model \cite{large_tan}. 

Since masses and couplings, thus branchings and cross-sections
 vary most rapidly with 
 m$_0$, m$_{1/2}$, it is natural to follow the commonly used way of 
presenting mSUGRA searches as a function of these two parameters for 
different fixed values of  tan$\beta$ and sign$(\mu$). The A$_0$ parameter
 is usually set to zero, since its variation has only a moderate
 effect on the results.

In Figs.1,2
 one can see total mSUGRA production cross-section (including 
associated $\tilde{\chi}\tilde{g}$, $\tilde{\chi}\tilde{q}$ and
 chargino-neutralino pair production)  as a function of 
m$_0$, m$_{1/2}$ for chosen sets of  tan$\beta$ and sign($\mu$).
The contribution of strongly interacting SUSY particles cross-sections
($\tilde{g}\tilde{g}$, $\tilde{g}\tilde{q}$, $\tilde{q}\tilde{q}$)   
is also shown separately by dashed lines.  
The jitter in the contours is caused by limited statistics. The total 
cross-section for the same values of m$_0$, m$_{1/2}$ but
 for different values of tan$\beta$ and sign($\mu$)
 differs slightly.
The bulk of the total cross-section for low values of m$_{1/2}$ comes from
$\tilde{g}\tilde{g}$, $\tilde{g}\tilde{q}$, $\tilde{q}\tilde{q}$, whereas
in the domains with extremely high masses of $\tilde{g}$, $\tilde{q}$  
the contribution of production of squarks or gluinos  associated
with much lighter (at the same m$_0$, m$_{1/2}$) charginos and neutralinos,
or even just chargino-neutralino pair production  may dominate.
 The shaded regions along the axes denote theoretically
(TH) and up to now experimentally (EX) excluded regions of 
parameter space as given in the ref.
\cite{ex_regions} .

All the results of this study are obtained with calculations made with
ISAJET 7.32, 7.44  generators \cite{ISAJET} and modified supplements therein.
\begin{myfig2}{hbtp}
\vspace*{0mm}
\hspace*{-5mm}\resizebox{.465\textwidth}{9cm}
                        {\includegraphics{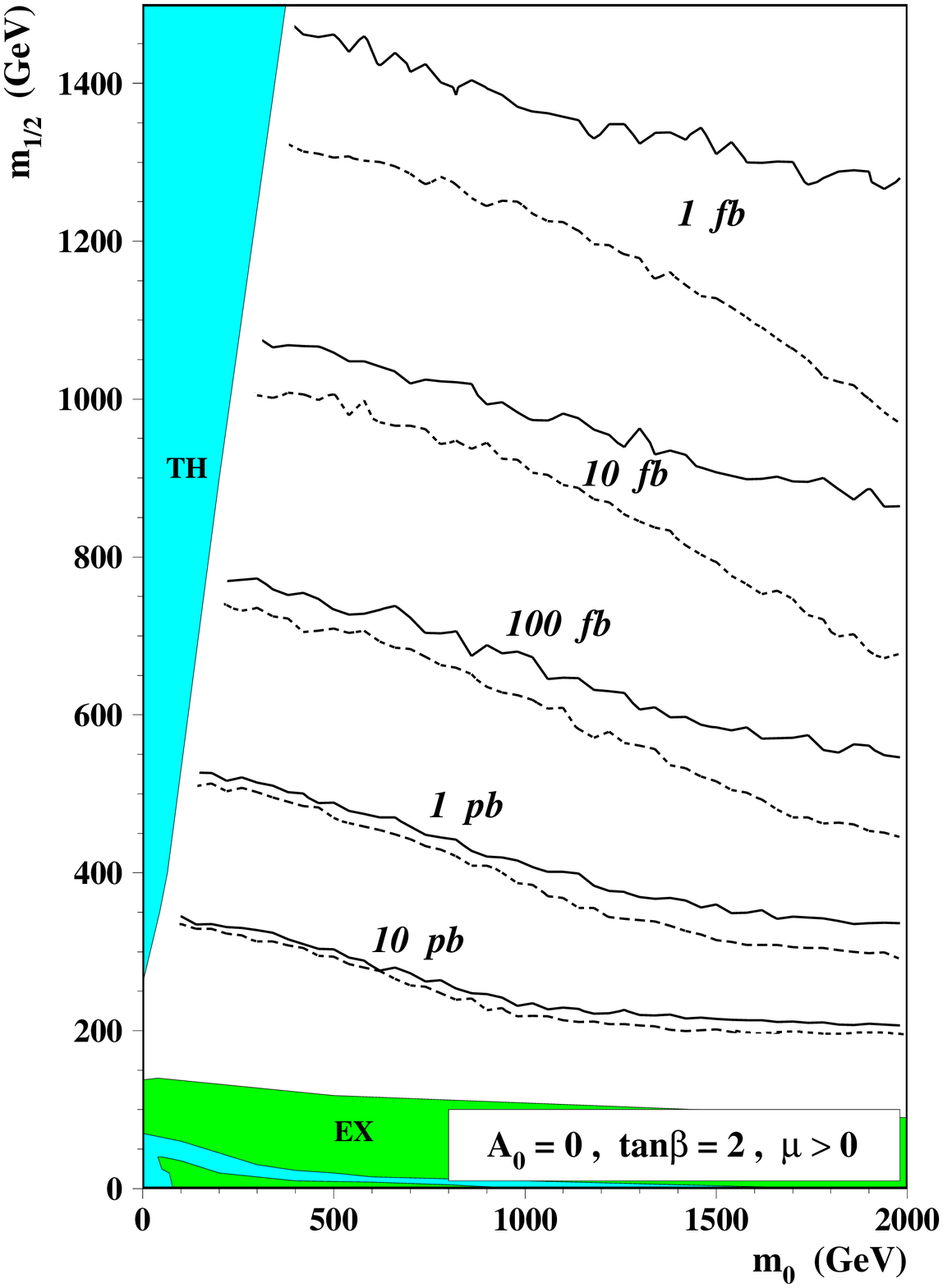}} & ~ &
\vspace*{0mm}
\hspace*{-5mm}\resizebox{.465\textwidth}{9cm}
                        {\includegraphics{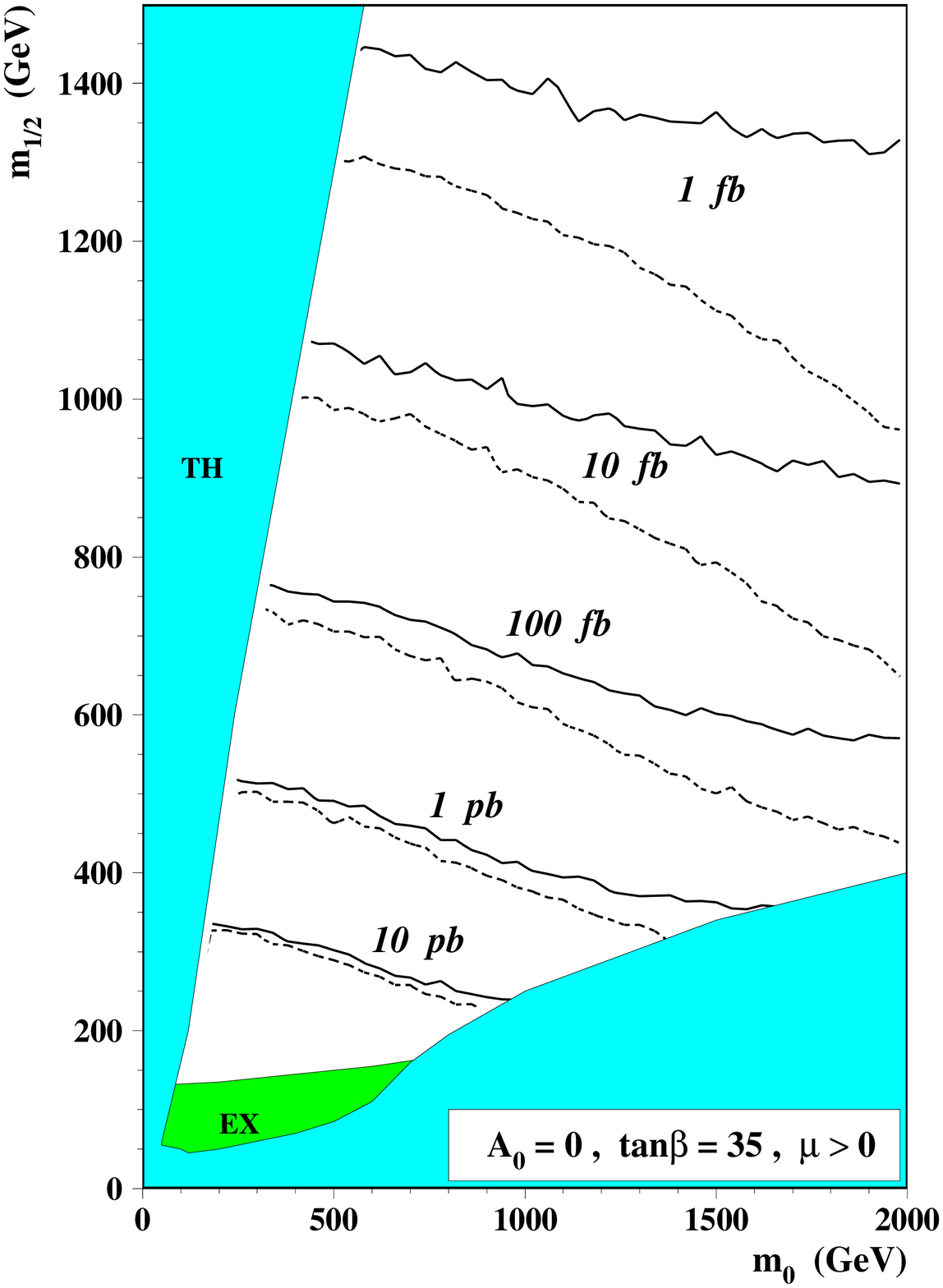}} \\
\vspace*{-5mm}
 \caption{ Total mSUGRA cross-section contours as a function of         
                       m$_0$, m$_{1/2}$  for A$_0$ = 0, tan$\beta$ = 2,
 $\mu$ $>$ 0 (solid line).
    The contribution
    of $\tilde{g}\tilde{g}$, $\tilde{g}\tilde{q}$, $\tilde{q}\tilde{q}$
    production alone is shown by dashed line.}    &~&
\vspace*{-5mm}
 \caption{ Same as Fig.1, except for tan$\beta$=35.}
\end{myfig2}
\vspace*{-10mm}

\section{ Difference in phenomenology between low and high values of 
tan$\beta$.}

Fig.3 a)
 shows the dependence of mass values of left selectron
 ($\tilde{e}_{L}$) and the lightest stau ($\tilde{\tau}_1$) as a function
 of tan$\beta$ at a representative point of mSUGRA parameter space. In
Fig.3 b)
 one sees the dramatic increase of the
 branching ratios of 
  lightest chargino ($\tilde{\chi}_{1}^{\pm}$, dashed line)
  and next-to-lightest neutralino ( $\tilde{\chi}_{2}^{0}$, solid line)
  into final states with lightest stau with increasing tan$\beta$, due to 
the increase of the tau Yukawa couplings.
This causes the enhancement of final states with taus at large 
 tan$\beta$. 
Similar examples of masses and branching ratios dependence on tan$\beta$
 for lower values of m$_0$ and m$_{1/2}$ (Tevatron range) can be found in 
ref. \cite{large_tan}.

Let us consider in more detail the taus, taking now for example the decay
 of a heavy
gluino in the appropriate domain of parameter space, and compare the
abundance of taus in the final states to see the difference between
large and small tan$\beta$. 
Figs.4 and 5
 show the decay schemes of heavy gluinos and squarks 
at high tan$\beta$, whilst 
Figs.6 and 7 
are for low tan$\beta$.
In Fig.4,
 for tan$\beta$=35,
 $\tilde{\chi}_{2}^{0}$ and
  $\tilde{\chi}_{1}^{\pm}$  branching ratios for 
decays into $\tilde{\tau}_{1,2}$ $\tau(\nu)$  exceed 60 $\%$.
To simplify the figure, similar intermediate states were grouped. For
instance, states 
 $\tilde{\chi}_{2}^{\pm}$Wbb and  $\tilde{\chi}_{2}^{\pm}$tb
were treated (summed up) as the same, though they have
different kinematics in principle (see the rightmost circular mark at
 $\tilde{\chi}_{2}^{\pm}$ horizontal line (1067 GeV).
 It is almost impossible to follow and 
calculate all the branchings in gluino decays;
 some small ones are not shown, thus resulting
in a small underestimate of the ``final'' states (at the level of  
 $\tilde{\chi}_{1}^{0}$) branching ratios.
The ten final states having the highest branching ratios 
are listed in the lower part of 
Figs.4 -- 7.
The right squarks ($\tilde{q}_R$) decay entirely
into  $\tilde{\chi}_{1}^{0}$ q  final state in the domain of mSUGRA 
parameter space where m$_{\tilde{q}_{R}}<$ m$_{\tilde{g}}$ as it is in
the point presented in
Figs.4 -- 7. 

In case of low  tan$\beta$=2,
 Fig.6,
 the decay chains of gluinos are not so complicated as in case
 of high tan$\beta$, Fig.4,
mainly due to the fact that 
m$_{\tilde{t}_1}<$ m$_{\tilde{\chi}_{2}^{\pm},\tilde{\chi}_{3,4}^{0}}$. 
 Right squarks again, as for high  tan$\beta$, decay entirely
into LSP + quarks. In addition,  at low tan$\beta$ 
the $\tilde{\tau}_{1,2}$ do not dominate in the decays of  
  $\tilde{\chi}_{1}^{\pm}$ and $\tilde{\chi}_{2}^{0}$, instead,
branchings of  $\tilde{\chi}_{1}^{\pm}$ and $\tilde{\chi}_{2}^{0}$ decays into
selectrons and smuons are enhanced. Thus final states of gluinos and 
squarks contain
more leptons (e, $\mu$) in case of low  tan$\beta$ than in case of high 
 tan$\beta$  in the chosen point of mSUGRA parameter space, as also 
discussed in \cite{lali}.

What is evident from the list of dominant decay modes at the bottom
 of Figs.4 -- 7 is that b's are abundantly produced and more so at large
tan$\beta$, but what is striking is the large number of final states containing
 $\tau$'s and multiple $\tau$'s at large  tan$\beta$,
providing in fact a method to constrain tan$\beta$ \cite{lali}.
Our aim here is to emphasize the instrumental implications of these 
observations.

\section{Differences between various domains of
 m$_{0}$, m$_{1/2}$ plane for high  tan$\beta$} 

Let us now consider 5 very different points in the 
(m$_{0}$,m$_{1/2}$) plane for fixed 
values of  tan$\beta$ = 35, $\mu$ $>$ 0  and  A$_0$=0 (see Table 1).
One can see that at the first point (130,240)
moderately above the Tevatron II reach, and at the second point (400,900),
this latter one already shown in
 Figs.3 -- 5, there is an abundant production of taus, whilst multiple b-jets
 originate
mainly from stop/sbottom production and partly from gluino (it is relatively
heavy to contribute significantly to the total cross-section).
Some difference in the mass values between Table 1 and Figs.3 -- 5 is due to 
the fact that values in Table 1 are obtained with ISAJET 7.44, whilst
those in Figs.3 -- 5 are calculated with  ISAJET 7.32.

\begin{table}[htb]
  \caption{Characteristic features of 5 mSUGRA points. High tan$\beta$ case .}
  \label{tab:1}
  \begin{center}
    \renewcommand{\arraystretch}{1.5}
    \setlength{\tabcolsep}{2mm}
\begin{tabular}{|c|c|c|c|c|c|} \hline
Mass of sparticles (GeV) &  \multicolumn{5}{c}{~~~ (m$_{0}$,m$_{1/2}$)
 values for  tan$\beta$ = 35, $\mu$ $>$ 0,  A$_0$=0 ~~~} \vline \\
 \cline{2-6} 
 or Branchings (\%) & 1 & 2 & 3 & 4 & 5 \\
 
 ~    & ~~ (130,240) ~~  & ~ (400,900) ~ & ~ (700,600) ~ &
   ~ (500,200) ~ & (1000,400) \\
\hline
$\tilde{g}$             & 598  & 2011 & 1407 & 537 & 1010 \\
$\tilde{u}_{L}$         & 541  & 1806 & 1396 & 661 & 1291 \\ 
$\tilde{t}_1$           & 390  & 1391 & 1018 & 429 &  846 \\
$\tilde{\chi}_{2}^{0}$  & 184  &  756 &  497 & 153 &  328 \\
$\tilde{e}_L$           & 220  &  741 &  818 & 522 & 1039 \\ 
$\tilde{\tau}_1$        &  99  &  442 &  641 & 440 &  892 \\ 
$\tilde{\chi}_{1}^{0}$  &  98  &  395 &  259 &  81 &  171 \\
\hline
\hline
Br($\tilde{g}$ $\rightarrow$ $\tilde{t}_1t$ + $\tilde{b}_1b$)   &
                    47.0 & 58.4 & 91.1 & 87.8 & 85.9$^*$ \\
Br($\tilde{q}_{R}$  $\rightarrow$ $\tilde{\chi}_{1}^{0}$ $q$) &
                   98.4 &  99.9 & 99.7 & 8 -- 25 & 6 -- 21 \\ 
Br($\tilde{q}_{R}$  $\rightarrow$ $\tilde{g}$ $q$) &
                    0   &  0  &  0    & 75 -- 92 & 79 -- 94 \\
Br($\tilde{\chi}_{2}^{0}$  $\rightarrow$  $\tilde{\tau}_{1,2}\tau$)  &
                    99.6 & 65.5 & 0$^{**}$ & 2.6$^{**}$   & 0$^{**}$ \\
Br($\tilde{\chi}_{1}^{\pm}$  $\rightarrow$  $\tilde{\tau}_{1,2}\nu$,
  $\tilde{\nu}\tau$) &
                    98.9 & 81.2 & 0$^{***}$  & 10.6$^{***}$  & 0$^{***}$ \\  
Br($\tilde{\chi}_{2}^{0}$  $\rightarrow$ $h$) {\bf $\cdot$} 
Br( $h$ $\rightarrow$ $b\bar{b}$) &
                    0  &  8  & 70 & 22.8$^{****}$ & 68 \\  
\hline
\hline
\end{tabular}
\end{center}
\hspace{3cm}
$ ^{*}$~~ Br($\tilde{g}$ $\rightarrow$ $\tilde{\chi}$ $t(b)t(b)$) \\

\hspace{3cm}
$ ^{**}$ Br($\tilde{\chi}_{2}^{0}$ $\rightarrow$
 $\tilde{\chi}_{1}^{0}\tau\tau$) \\

\hspace{3cm}
$ ^{***}$ Br($\tilde{\chi}_{1}^{\pm}$ $\rightarrow$
 $\tilde{\chi}_{1}^{0}\tau\nu$) \\

\hspace{3cm}
$ ^{****}$ Br($\tilde{\chi}_{2}^{0}$ $\rightarrow$
 $\tilde{\chi}_{1}^{0}b\bar{b}$) \\
\end{table}

The third point (700,600) is characterized by an increase 
of b-production due to production of
 not so heavy gluino, but mainly from $\tilde{\chi}_{2}^{0}$.
At the same time one can see a drastic decrease of tau production at the third
parameter space point, the remaining source of
them being the decay of W-bosons and, to some extent, decays of B-mesons. 

At the forth point with relatively light gluino and squarks 
(slightly above the Tevatron II reach, as the first point) 
$\tau$ production is moderate due to smaller branchings of
 $\tilde{\chi}_{2}^{0}$ and
 $\tilde{\chi}_{1}^{\pm}$, whilst b-production is significant (comparable
to that at the previous point),  despite
$\tilde{\chi}_{2}^{0}$ 
$\rightarrow$  $\tilde{\chi}_{1}^{0}$ $h$ $\rightarrow$  $\tilde{\chi}_{1}^{0}$
$b\bar{b}$ being kinematically forbidden,
but there is a large  $\tilde{\chi}_{2}^{0}$ 
 3-body decay with   
Br($\tilde{\chi}_{2}^{0}$  $\rightarrow$ $\tilde{\chi}_{1}^{0}b\bar{b}$) 
$\approx$ 23 \%.
 This is due to the fact that at this point 
 the gluino mass is smaller than masses of
squarks (except $\tilde{t}_1$) and branching ratios of squark decays into
gluino + quark are significant, 
 even right squarks now
decay mostly through gluino + quark. In turn, the
Br($\tilde{g}$ $\rightarrow$ $\tilde{b}_1b$) $\approx$ 88 \% at this point.
 
At the fifth point (1000,400) again the gluino mass is smaller than masses of
squarks (except $\tilde{t}_1$) and squarks mostly
decay through gluino + quark. In addition, 
$\tilde{\chi}_{2}^{0}$ 
$\rightarrow$  $\tilde{\chi}_{1}^{0}$ $h$ $\rightarrow$  $\tilde{\chi}_{1}^{0}$
$b\bar{b}$ is kinematically allowed, though the yield of 
 $\tilde{\chi}_{2}^{0}$ is lower than at point 3  because left squarks 
often decay via gluino + quark , and Br($\tilde{q}_{L}$ $\rightarrow$ 
 $\tilde{\chi}_{2}^{0}$ $q$) is only $\approx$ 17 \%  against  $\approx$ 32 \%
at point 3.
It results in a significant
b-production, since branching of gluino decay into top/bottom final state
remains high ($\geq$ 80 \%). It also leads to a slight
increase of tau production at this point (via $t$ $\rightarrow$ $Wb$  
 $\rightarrow$ $\tau \nu b$). To conclude this brief overview,
b-jets are abundantly produced over most of the m$_{0}$, m$_{1/2}$ plane,
i.e. from the Tevatron reach up to the maximal LHC  $\tilde{g}$, $\tilde{q}$
mass reach,
whilst $\tau$'s are abundant in the region 
m$_{1/2}$ $>$ (1 -- 1.5)$\cdot$m$_0$.

To give a feeling about the situation for ``intermediate'' tan$\beta$,
we show in Table 2 the masses and branching ratios for the same 5 
(m$_{0}$,m$_{1/2}$) points as in Table 1, but for tan$\beta$=10. 
The only essential difference between the two tables 
(two values of  tan$\beta$) is in the lower branchings of 
$\tilde{\chi}_{2}^{0}$, $\tilde{\chi}_{1}^{\pm}$ decays into
 $\tilde{\tau}_{1,2}\tau$ and $\tilde{\tau}_{1,2}\nu$ respectively for
 tan$\beta$=10.

\begin{table}[htb]
  \caption{Characteristic features of 5 mSUGRA points.
 Intermediate tan$\beta$ case .}
  \label{tab:2}
  \begin{center}
    \renewcommand{\arraystretch}{1.5}
    \setlength{\tabcolsep}{2mm}
\begin{tabular}{|c|c|c|c|c|c|} \hline
Mass of sparticles (GeV) &  \multicolumn{5}{c}{~~~ (m$_{0}$,m$_{1/2}$)
 values for  tan$\beta$ = 10, $\mu$ $>$ 0,  A$_0$=0 ~~~} \vline \\
 \cline{2-6} 
 or Branchings (\%) & 1 & 2 & 3 & 4 & 5 \\
 
 ~    & ~~ (130,240) ~~  & ~ (400,900) ~ & ~ (700,600) ~ &
   ~ (500,200) ~&  (1000,400) \\
\hline
$\tilde{g}$             & 604  & 2015 & 1426 & 537 & 1012 \\
$\tilde{u}_{L}$         & 544  & 1808 & 1393 & 661 & 1291 \\ 
$\tilde{t}_1$           & 388  & 1386 & 1016 & 429 &  844 \\
$\tilde{\chi}_{2}^{0}$  & 182  &  755 &  498 & 153 &  328 \\
$\tilde{e}_L$           & 220  &  741 &  816 & 522 & 1039 \\ 
$\tilde{\tau}_1$        & 159  &  522 &  730 & 502 & 1003 \\ 
$\tilde{\chi}_{1}^{0}$  &  97  &  395 &  259 &  80 &  170 \\
\hline
\hline
Br($\tilde{g}$ $\rightarrow$ $\tilde{t}_1t$ + $\tilde{b}_1b$)   &
                    37.0 & 53.5 & 80.0 & 34.0$^*$ & 84.0$^*$ \\
Br($\tilde{q}_{R}$  $\rightarrow$ $\tilde{\chi}_{1}^{0}$ $q$) &
                   98.2 &  99.9 & 99.7 & 8 -- 25 & 6 -- 20 \\ 
Br($\tilde{q}_{R}$  $\rightarrow$ $\tilde{g}$ $q$) &
                    0   &  0  &  0  & 75 -- 92 & 78 -- 93 \\
Br($\tilde{\chi}_{2}^{0}$  $\rightarrow$  $\tilde{\tau}_{1,2}\tau$) &
                    84.8 & 21.0 & 0$^{**}$ & 2.1$^{**}$ & 0$^{**}$ \\
Br($\tilde{\chi}_{1}^{\pm}$  $\rightarrow$  $\tilde{\tau}_{1,2}\nu$,
  $\tilde{\nu}\tau$) &
                    68.3 & 32.9 & 0$^{***}$  & 10.7$^{***}$ & 0$^{***}$ \\  
Br($\tilde{\chi}_{2}^{0}$  $\rightarrow$ $h$) {\bf $\cdot$} 
Br( $h$ $\rightarrow$ $b\bar{b}$) &
                    0  &  26.0 & 74.0 & 18.6$^{****}$ & 74.0 \\  
\hline
\hline
\end{tabular}
\end{center}
\hspace{3cm}
$ ^{*}$~~ Br($\tilde{g}$ $\rightarrow$ $\tilde{\chi}$ $t(b)t(b)$) \\

\hspace{3cm}
$ ^{**}$ Br($\tilde{\chi}_{2}^{0}$ $\rightarrow$
 $\tilde{\chi}_{1}^{0}\tau\tau$) \\

\hspace{3cm}
$ ^{***}$ Br($\tilde{\chi}_{1}^{\pm}$ $\rightarrow$
 $\tilde{\chi}_{1}^{0}\tau\nu$) \\

\hspace{3cm}
$ ^{****}$ Br($\tilde{\chi}_{2}^{0}$ $\rightarrow$
 $\tilde{\chi}_{1}^{0}b\bar{b}$) \\
\end{table}

\section{Generalization for MSSM}

The considerations of the previous section can be extended to the
case of more the more general MSSM model.
 One can deduce from Table 1 that if
 m$_{\tilde{\chi}_{2}^{0},\tilde{\chi}_{1}^{\pm}}$ $>$ m$_{\tilde{\tau}_{1}}$
and tan$\beta$ is high enough,
one can expect an abundant production  of taus in the MSSM final states,
 as in points (130,240) and (400,900) in Table 1.
This is illustrated by
 MSSM points 1 and 2 in Table 3.

Then, if $\tilde{\chi}_{2}^{0}$ $\rightarrow$ $\tilde{\chi}_{1}^{0}$ $h$
 decay is kinematically allowed and production of
  $\tilde{\chi}_{2}^{0}$ is significant
(both in squark-gluino cascades or just in chargino-neutralino pair
production, the latter has important effect if 
m$_{\tilde{\chi}_{2}^{0},\tilde{\chi}_{1}^{\pm}}$ $\ll$ 
min(m$_{\tilde{g}}$,
m$_{\tilde{q}}$) and corresponds to the high values of m$_{1/2}$ in mSUGRA),
then it results in a noticeable production of b-jets coming
from lightest Higgs decay into $b\bar{b}$, see point 3 in Tables 1 -- 3,
generalizing in MSSM  the observations done in \cite{hbb}.

Finally,
 if m$_{\tilde{q}}$ $>$  m$_{\tilde{g}}$, it enhances mostly  the
 yield of b-quarks, despite the known decrease of $\tilde{\chi}_{2}^{0}$
production (with a corresponding decrease of the $h$ $\rightarrow$ $b\bar{b}$
as a source of b-jets) due to the fact that squarks decay through
 gluino + quarks
and the source of  $\tilde{\chi}_{2}^{0}$ from
$\tilde{q}$ $\rightarrow$  $\tilde{\chi}_{2}^{0}$ $q$ is reduced.
In this case b-jets are mainly produced in the decays of gluinos, which are,
in turn, produced by both left and right squarks.
One can compare
the points 4 and 5  in Tables 1 -- 3.

For the particular choice of MSSM parameters in Table 3, the 
Br($\tilde{g}$ $\rightarrow$ $\tilde{\chi}$ $t(b)t(b)$)
 in the MSSM case
 is smaller than that of a similar point of mSUGRA, due to
the fact that in mSUGRA $\tilde{t}_1$ and  $\tilde{b}_1$ are much lighter
than all other squarks, whereas in MSSM point 5 only  $\tilde{b}_1$ is 
a bit lighter(1280 GeV) than other squarks ($\sim$ 1300 GeV),
 but  $\tilde{b}_2$,$\tilde{t}_{1,2}$ are
heavier (1320, 1307 and 1309 GeV respectively). 
But if in MSSM the third generation squarks are, as in mSUGRA, significantly 
lighter than first two generation squarks,
 it immediately results in the increase
of Br($\tilde{g}$ $\rightarrow$ $\tilde{\chi}$ $t(b)t(b)$) up to values of
80-90 \%, as in mSUGRA.  
In the MSSM domain explored so far,
masses of charginos and neutralinos are obtained still 
assuming universal gaugino
mass at the GUT scale. The common slepton masses are taken fairly arbitrary
in such a way that m$_{\tilde{\chi}_{1}^{0}}$ $<$ m$_{\tilde{\tau}_1}$ $<$
m$_{\tilde{\chi}_{2}^{0}}$.
 
\begin{table}[htb]
  \caption{Characteristic features of 5 MSSM points.}
  \label{tab:3}
  \begin{center}
    \renewcommand{\arraystretch}{1.5}
    \setlength{\tabcolsep}{2mm}
\begin{tabular}{|c|c|c|c|c|c|} \hline
Mass of sparticles (GeV) &  \multicolumn{5}{c}{~~
tan$\beta$ = 35, $\mu$ = 500 GeV, m$_A$ = 1000 GeV,  A$_t$=A$_b$=A$_{\tau}$=0  
~~} \vline \\
 \cline{2-6} 
or Branchings (\%) & \hspace{8mm} 1 \hspace{8mm} 
                   & \hspace{8mm} 2 \hspace{8mm}
                   & \hspace{8mm} 3 \hspace{8mm}
                   & \hspace{8mm} 4 \hspace{8mm} 
                   & \hspace{8mm} 5 \hspace{8mm} \\
\hline
$\tilde{g}$             & 600  & 2000 & 1400 & 500 & 1000 \\
$\tilde{q}$             & 500  & 1800 & 1300 & 650 & 1300 \\ 
$\tilde{\chi}_{2}^{0}$  & 177  &  485 &  407 & 135 &  279 \\
$\tilde{e}_L$           & 200  &  400 &  700 & 650 & 1300 \\ 
$\tilde{\tau}_1$        & 107  &  362 &  679 & 628 & 1288 \\ 
$\tilde{\chi}_{1}^{0}$  &  91  &  319 &  218 &  69 &  145 \\
\hline
\hline
Br($\tilde{g}$ $\rightarrow$ $\tilde{t}_1t$ + $\tilde{b}_1b$)   &
                    23.3 & 31.8 & 21.6 & 17.8$^{*}$ &  36.5$^{*}$     \\
Br($\tilde{q}_{R}$  $\rightarrow$ $\tilde{\chi}_{1}^{0}$ $q$) &
                    99.9 & 96.8 & 98.7 & 5 -- 17 &5 -- 18  \\ 
Br($\tilde{q}_{R}$  $\rightarrow$ $\tilde{g}$ $q$) &
                    0   &  0  &  0  & 83 -- 95 & 81 -- 94 \\
Br($\tilde{\chi}_{2}^{0}$  $\rightarrow$  $\tilde{\tau}_{1,2}\tau$)  &
                    100.0 & 40.5 & 0$^{**}$ & 1.8$^{**}$ & 0$^{**}$ \\
Br($\tilde{\chi}_{1}^{\pm}$  $\rightarrow$  $\tilde{\tau}_{1,2}\nu$,
  $\tilde{\nu}\tau$) &
                     99.8 & 45.3 & 0$^{***}$ & 10.0$^{***}$  & 0$^{***}$ \\  
Br($\tilde{\chi}_{2}^{0}$  $\rightarrow$ $h$) {\bf $\cdot$} 
Br( $h$ $\rightarrow$ $b\bar{b}$)       & 
                    0   & 26.1& 72.0 & 21.8$^{****}$ & 65.9 \\  
\hline
\hline
\end{tabular}
\end{center}
\hspace{3cm}
$ ^{*}$~~ Br($\tilde{g}$ $\rightarrow$ $\tilde{\chi}$ $t(b)t(b)$) \\

\hspace{3cm}
$ ^{**}$ Br($\tilde{\chi}_{2}^{0}$ $\rightarrow$
 $\tilde{\chi}_{1}^{0}\tau\tau$) \\

\hspace{3cm}
$ ^{***}$ Br($\tilde{\chi}_{1}^{\pm}$ $\rightarrow$
 $\tilde{\chi}_{1}^{0}\tau\nu$) \\

\hspace{3cm}
$ ^{****}$ Br($\tilde{\chi}_{2}^{0}$ $\rightarrow$
 $\tilde{\chi}_{1}^{0}b\bar{b}$) \\
\end{table}

\section{Results and conclusions}

Our findings are summarized in
Figs.8 -- 10
where  we plot the probabilities to find at least 1, 3 or 5 b-jets 
{\it per event}
respectively (with E$_T^b$ $>$ 50 GeV in $| \eta^b |$ $<$ 2.4) 
over mSUGRA parameter space for various sets of tan$\beta$, sign($\mu$).
Even the probabilities for 
{\it 3 b-jets per event}
 within CMS acceptance
(and with E$_T^b$ $>$ 50 GeV !) are significant for all  tan$\beta$
over most of parameter
space, from a few \% to $\sim$ 40 \%, the region m$_0$ $>$ m$_{1/2}$ being
the richest one. 
One can see that the effect of b Yukawa couplings increase  with tan$\beta$ 
is visible only in
 Fig.10,
i.e. asking at least 5 b-jets per event.
We would like to emphasize that the exceptionally large probability per event
 to find a b-jet in a SUSY event is not limited  to large m$_{0}$, m$_{1/2}$
(i.e. large $\tilde{g}$, $\tilde{q}$ values), but
{\it 
is valid throughout the whole plane, i.e. even for $\tilde{g}$, $\tilde{q}$ 
 masses $\sim$ 500 GeV,  just above the Tevatron reach,
} as illustrated by columns 1 and 4 in Tables 1,2.

The same probability distributions for at least 1, 2 and 3 taus
{\it per event} 
are shown in Figs.11 -- 13
 respectively (again for E$_T^{\tau}$ $>$ 50 GeV in $| \eta^{\tau} |$ $<$ 2.4).
 Here one can see a much more pronounced difference between 
various values of tan$\beta$ already with at least one tau in the final state
(Fig.11).
The  clearly visible ridge on the m$_{0}$, m$_{1/2}$ plane corresponds to
the mSUGRA domains of parameter space where $\tilde{\chi}_{2}^{0}$ and
$\tilde{\chi}_{1}^{\pm}$ decay into sleptons. The ridge becomes more and more
pronounced as tan$\beta$ increases thanks to increasing  $\tau$ Yukawa 
couplings and decreasing  $\tau$ mass relative
to $\tilde{e}$ and $\tilde{\mu}$. 
 One can also notice a significant (15 -- 20 \%) level
 of tau production throughout the  m$_{0}$, m$_{1/2}$ plane, mainly from
W and b-jets. 

It is also clear from the comparison of
 Figs.8 -- 10 with
 Figs.11 -- 13 
that the domains of 
parameter space where taus and b-jets are produced abundantly are 
complementary, but together cover most of parameter space ! 
It will thus be of utmost importance to have a high performance microvertex 
device to disentangle the complicated final states (Figs.4 -- 7) through
b and $\tau$ tagging, and the more so the higher is tan$\beta$; 
whether  tan$\beta$ is large
 can be detected already in early LHC running through the 
modifications of the expected shape  of opposite sign dilepton mass spectra 
discussed in ref \cite{lali} and 
due precisely to the increased $\tau$ production.
In general, the importance of the b-tagging capability of CMS
 increases with the 
SUSY mass scale whatever tan$\beta$ due to complexification of the 
final states with increasing masses,
 but it is already of crucial importance in the mass range just beyond the
Tevatron II reach, m$_{\tilde{g}}$, m$_{\tilde{g}}$ $\gtrsim$ 400 GeV, i.e,
m$_{0}$ $\gtrsim$ 200 GeV, m$_{1/2}$ $\gtrsim$ 400 GeV  in this model
(Tables 1 -- 3 and Figs.8 -- 10).

The main conclusions of our study are the following.

It has been known  quantitatively
for a long time that b-tagging will be important for 
SUSY studies. 
Our investigations further strengthen this feeling.
Within the mSUGRA model we find surprisingly high probabilities
per event,
in the few tens of \%, to find hard b-jets within CMS acceptance,
 even for multiple-b final states, the effect increasing
moderately  with tan$\beta$. 
What is more spectacular and not so well known and appreciated,
is the rapid increase of $\tau$-probability
 and multi-$\tau$ probability per event with increasing  tan$\beta$.
Furthermore, 
there is an almost complete complementarity 
in parameter space coverage by those
two experimental signatures. This implies that
a high performance microvertex detector will be essential for SUSY event
analysis, especially if tan$\beta$ $\geq$ 10, and the more so more
massive are the squarks and gluinos.
 Every effort should thus be made to have a third pixel layer in the
 CMS  tracker to have redundancy and a safety margin to ensure 
 in the impact parameter measurements.
At least provisions must be made in the inner tracker mechanical design
and beam pipe design that this third layer can be inserted if SUSY
shows up at the LHC rendes-vous. 
 Also every effort should be made to keep the innermost
pixel layer at 4 cm radius even in high luminosity conditions,
where probably frequent replacements will be needed,
as it improves significantly the b-tagging performance of CMS
as investigations in ref.
  \cite{tracker_TDR,caner} have shown.
It is also only with high luminosity that the highest masses are attainable,
 and it is for these masses that the decays are most complicated and b-tagging
 most useful.

An objection can be raised to these conclusions, namely that they are based
 on a very specific SUSY scenario - mSUGRA.
However, these conclusions will be generally true in whatever SUSY
model as long as $\tilde{t}_1$ or $\tilde{b}_1$ are substantially lighter 
than the remaining squarks, and the $\tilde{\tau}$ is lighter than 
$\tilde{\chi}_{2}^{0}$ and/or $\tilde{\chi}_{1}^{\pm}$
and $\tilde{\tau}_1$ is not significantly heavier than other sleptons,
 as was discussed in section 5.
Some preliminary results in the
more general MSSM context have already been shown,
and those investigations are being actively pursued at present.


\begin{figure}[hbtp]
\begin{center}
\vspace*{-5mm}
\hspace*{0mm}\resizebox{12cm}{!}
                        {\includegraphics{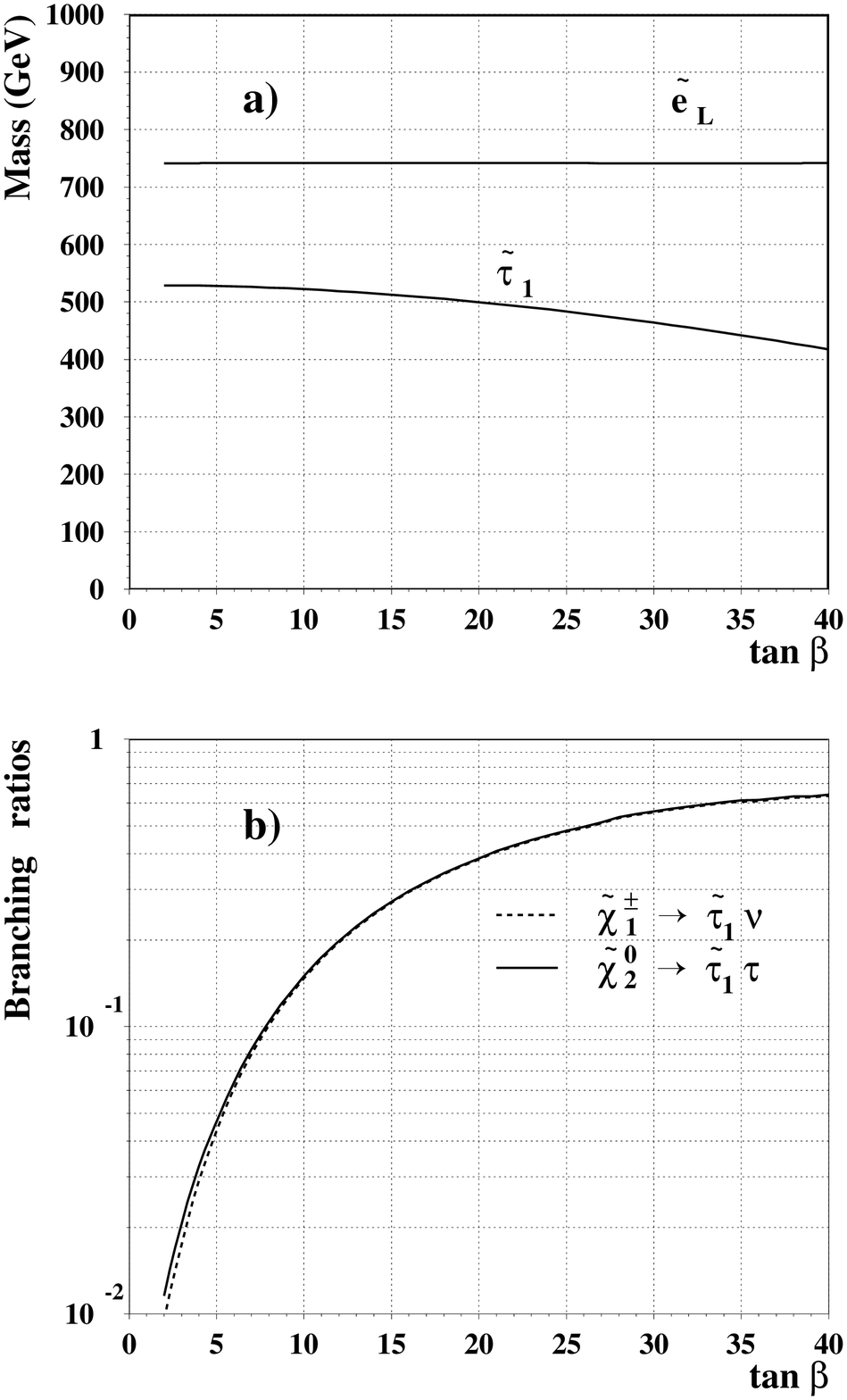}}
\vspace*{0mm}
 \caption{a) Masses of left selectron ($\tilde{e}_{L}$) and lightest stau
  ($\tilde{\tau}_1$) vs tan$\beta$, b) branching ratios of 
  lightest chargino ($\tilde{\chi}_{1}^{\pm}$, dashed line)
  and next-to-lightest neutralino ( $\tilde{\chi}_{2}^{0}$, solid line)
  into final states with lightest stau as a function of  tan$\beta$
  for m$_0$=400 GeV, m$_{1/2}$=900 GeV, A$_0$=0 and $\mu$ $>$ 0.}
\end{center}
\end{figure}
\begin{figure}[hbtp]
\begin{center}
\vspace*{0mm}
\hspace*{0mm}\resizebox{0.99\textwidth}{21cm}
                        {\includegraphics{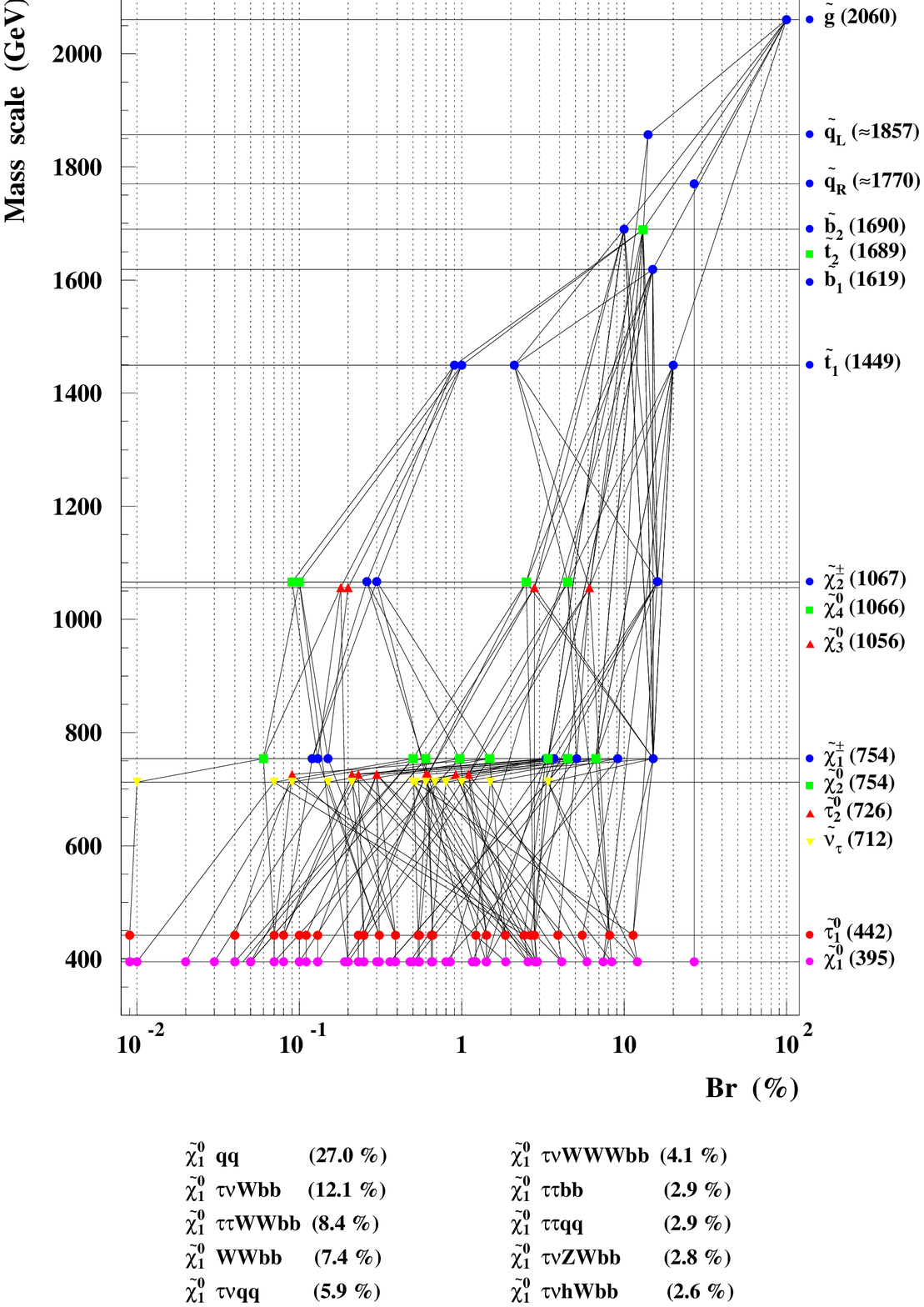}}
 \caption{ Typical decay modes for massive (2060 GeV) gluino
 {\bf for high tan$\beta$ = 35} (m$_0$ = 400 GeV, m$_{1/2}$ = 900 GeV,
 A$_0$ = 0 and $\mu$ $>$ 0). The ten largest individual decay modes
 are specified. Note multiplicity of b and $\tau$ final states.
}
\end{center}
\end{figure}
\begin{figure}[hbtp]
\begin{center}
\vspace*{0mm}
\hspace*{0mm}\resizebox{0.99\textwidth}{21cm}
                        {\includegraphics{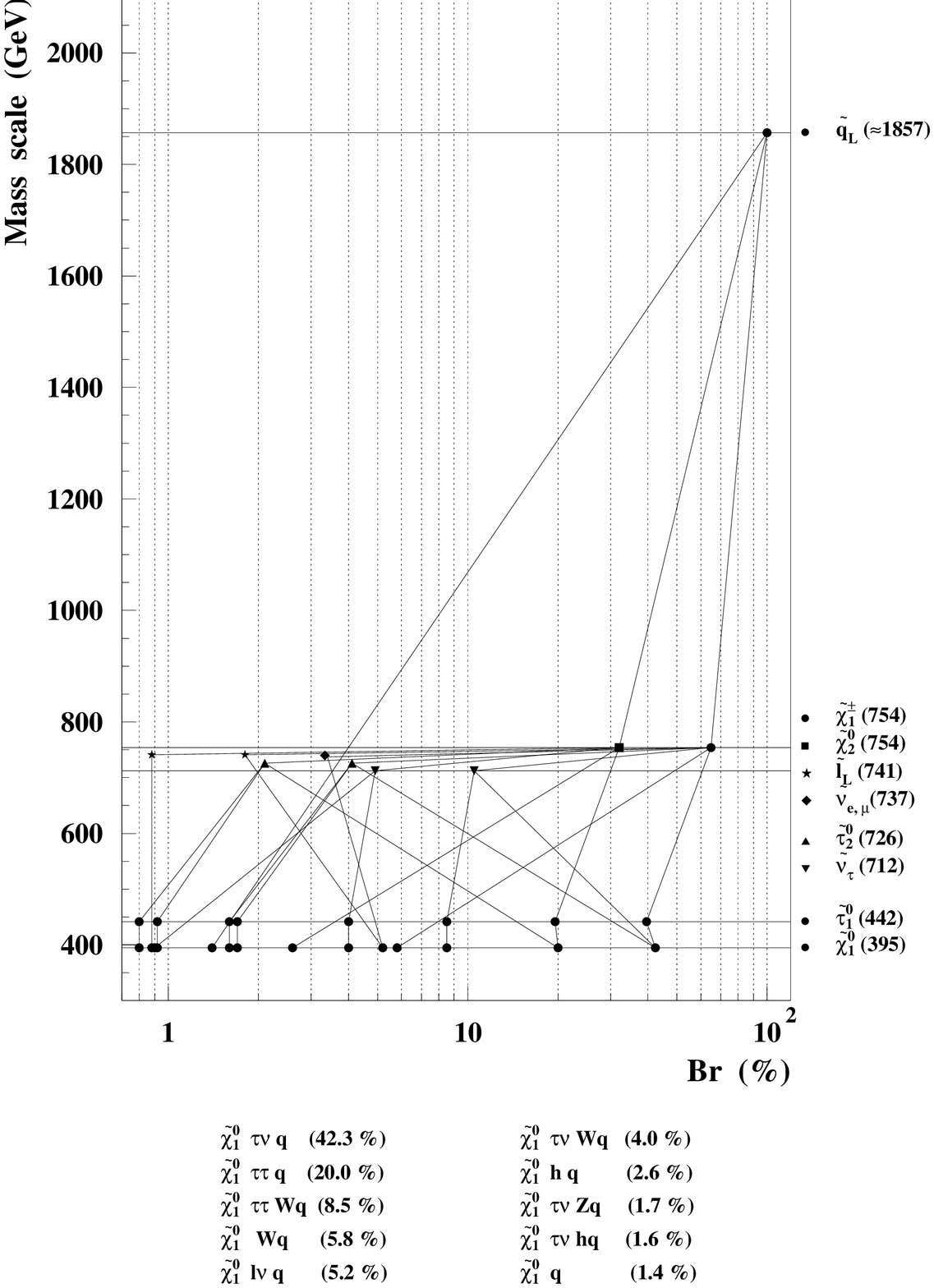}}
 \caption{ Typical decay modes for massive (1857 GeV) left squark
 {\bf for high tan$\beta$ = 35} (m$_0$ = 400 GeV, m$_{1/2}$ = 900 GeV,
 A$_0$ = 0 and $\mu$ $>$ 0). 
}
\end{center}
\end{figure}
\begin{figure}[hbtp]
\begin{center}
\vspace*{0mm}
\hspace*{0mm}\resizebox{0.99\textwidth}{21cm}
                        {\includegraphics{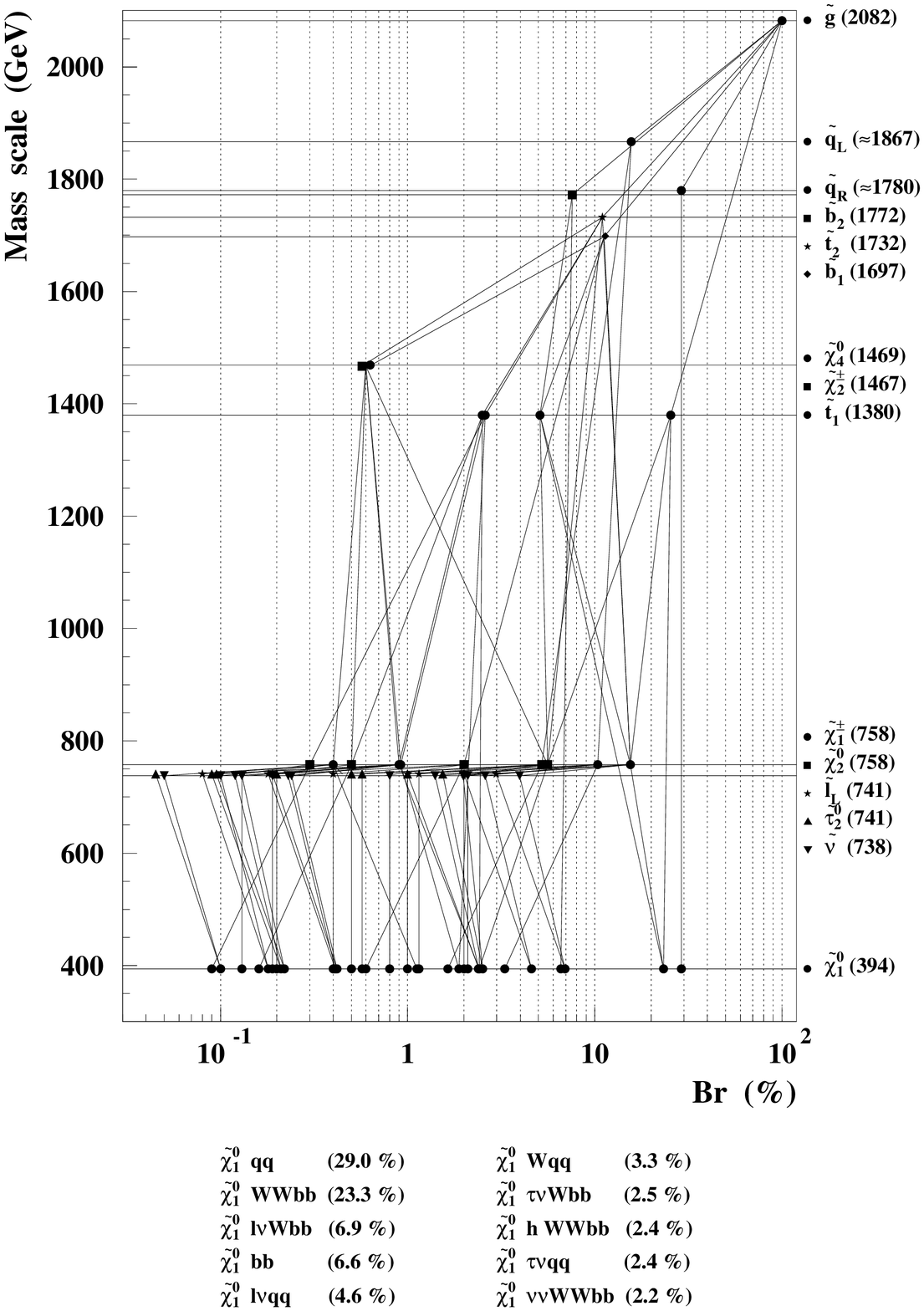}}
 \caption{ Typical decay modes for massive (2082 GeV) gluino
 {\bf for low tan$\beta$ = 2} (m$_0$ = 400 GeV, m$_{1/2}$ = 900 GeV,
 A$_0$ = 0 and $\mu$ $>$ 0). Note again large number of final states with b's.
}
\end{center}
\end{figure}
\begin{figure}[hbtp]
\begin{center}
\vspace*{0mm}
\hspace*{0mm}\resizebox{0.99\textwidth}{21cm}
                        {\includegraphics{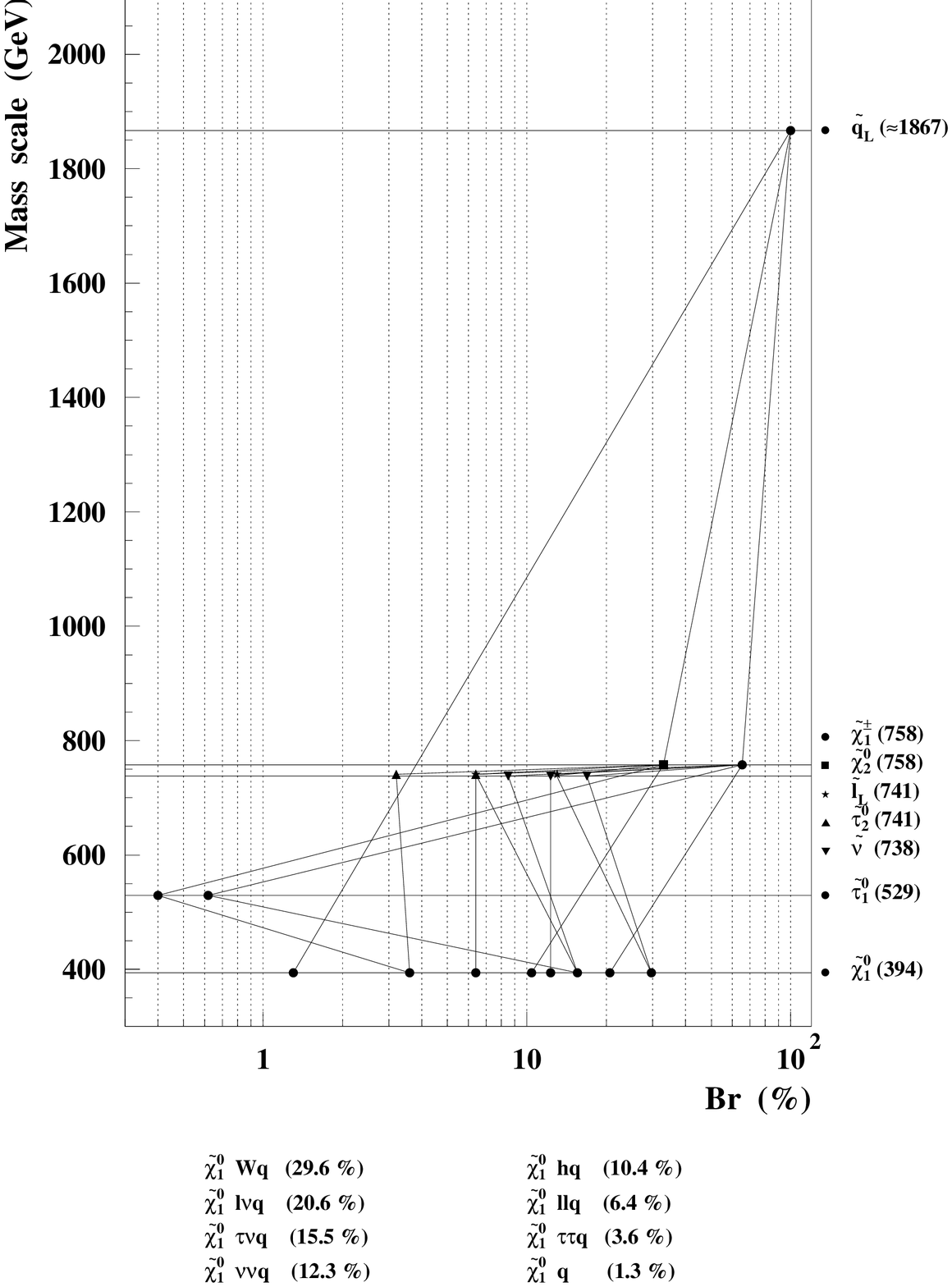}}
 \caption{ Typical decay modes for massive (1867 GeV) left squark
 {\bf for low tan$\beta$ = 2} (m$_0$ = 400 GeV, m$_{1/2}$ = 900 GeV,
 A$_0$ = 0 and $\mu$ $>$ 0).
}
\end{center}
\end{figure}
\begin{figure}[hbtp]
\begin{center}
\vspace*{0mm}
\hspace*{0mm}\resizebox{0.99\textwidth}{22cm}
                        {\includegraphics{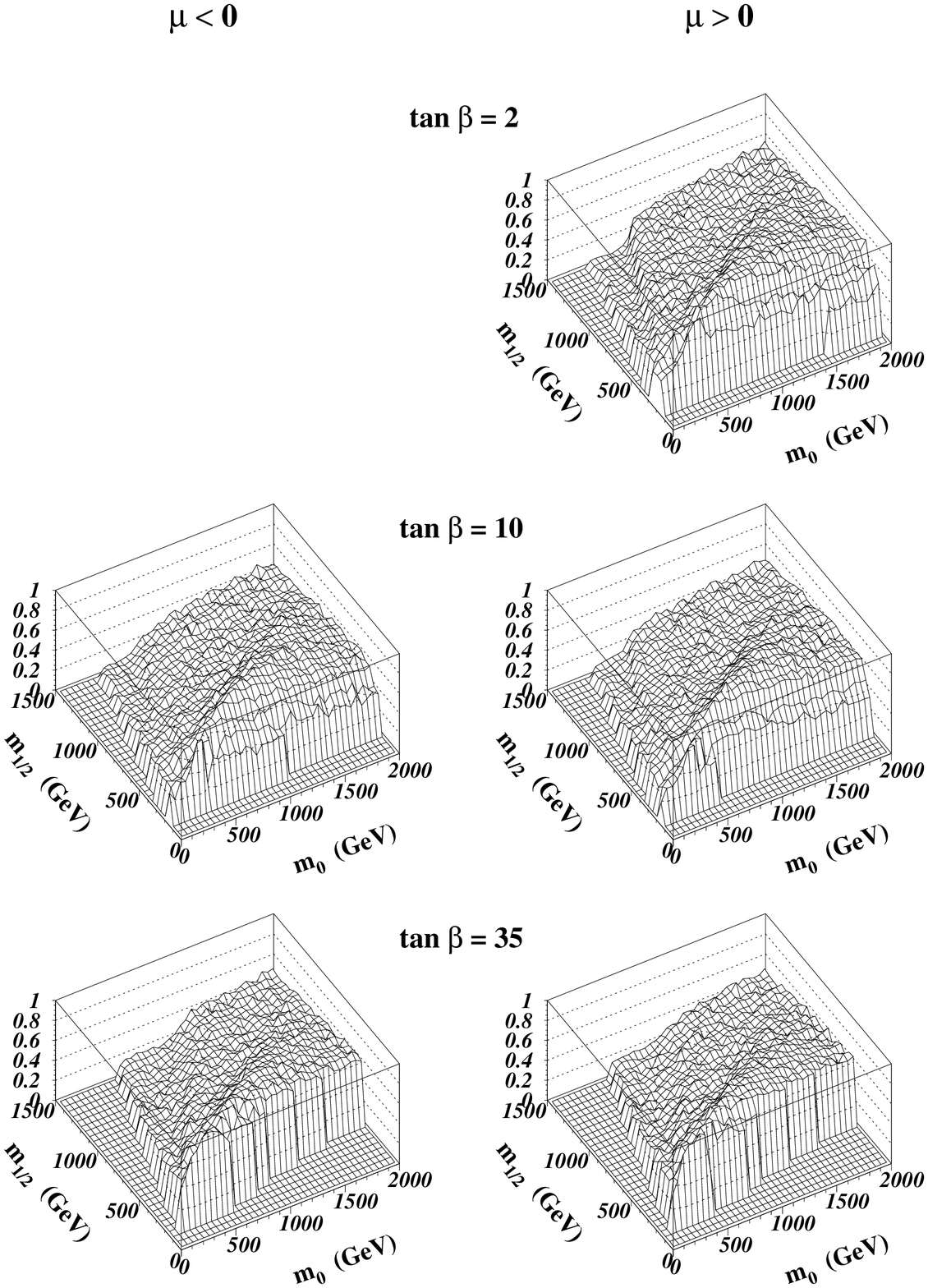}}
 \caption{ Probability to find at least one b-jet per event 
  with E$_T^b >$ 50 GeV in CMS acceptance 
  $| \eta^b |$ $<$ 2.4 for various mSUGRA parameter domains.
Notice large probability for all tan$\beta$ and $\mu$ and 
over all explorable m$_0$, m$_{1/2}$ domain.  
}
\end{center}
\end{figure}
\begin{figure}[hbtp]
\begin{center}
\vspace*{0mm}
\hspace*{0mm}\resizebox{0.99\textwidth}{22cm}
                        {\includegraphics{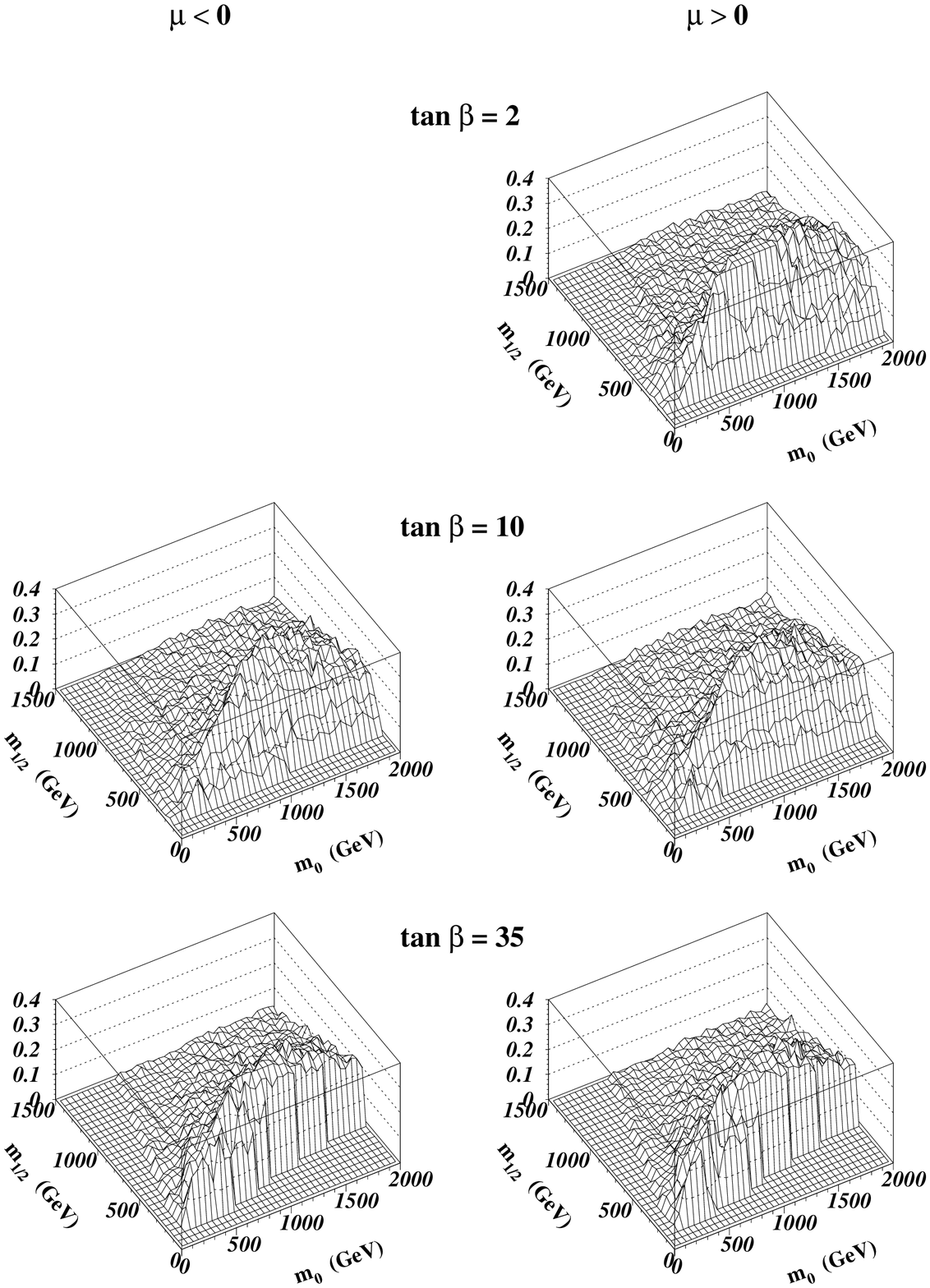}}
 \caption{ Same as in Fig.8, but for at least 3 b-jets.}
\end{center}
\end{figure}
\begin{figure}[hbtp]
\begin{center}
\vspace*{0mm}
\hspace*{0mm}\resizebox{0.99\textwidth}{22cm}
                        {\includegraphics{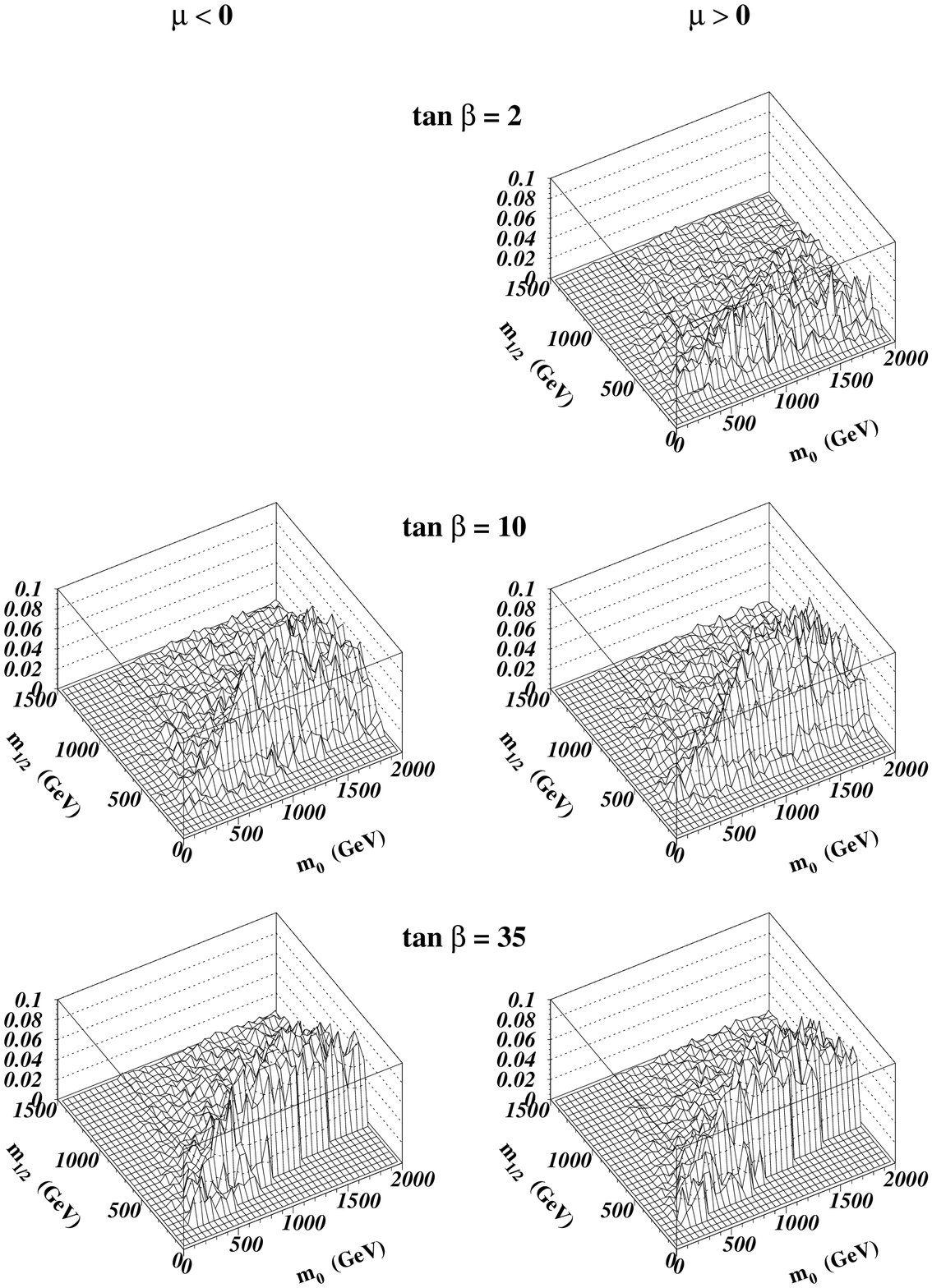}}
 \caption{ Same as in Fig.8, but for at least 5 b-jets.}
\end{center}
\end{figure}
\begin{figure}[hbtp]
\begin{center}
\vspace*{0mm}
\hspace*{0mm}\resizebox{0.99\textwidth}{22cm}
                        {\includegraphics{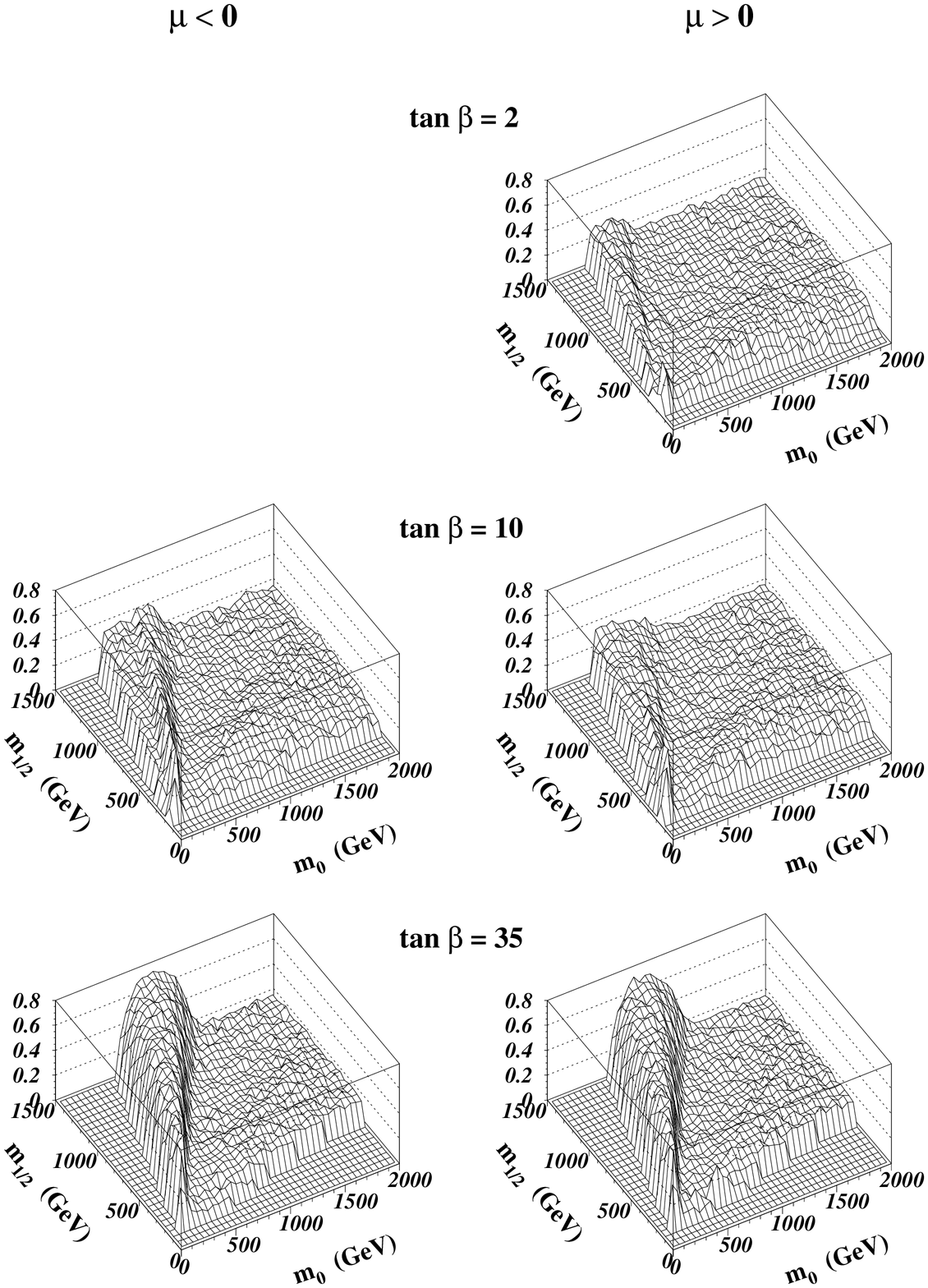}}
 \caption{ Probability to find at least one tau per event
   with E$_T^{\tau} >$ 50 GeV in CMS acceptance
  $| \eta^{\tau} |$ $<$ 2.4 for various mSUGRA parameter domains.
Note large increase with increasing tan$\beta$, but significant
probability nonetheless throughout parameter space.
}
\end{center}
\end{figure}
\begin{figure}[hbtp]
\begin{center}
\vspace*{0mm}
\hspace*{0mm}\resizebox{0.99\textwidth}{22cm}
                        {\includegraphics{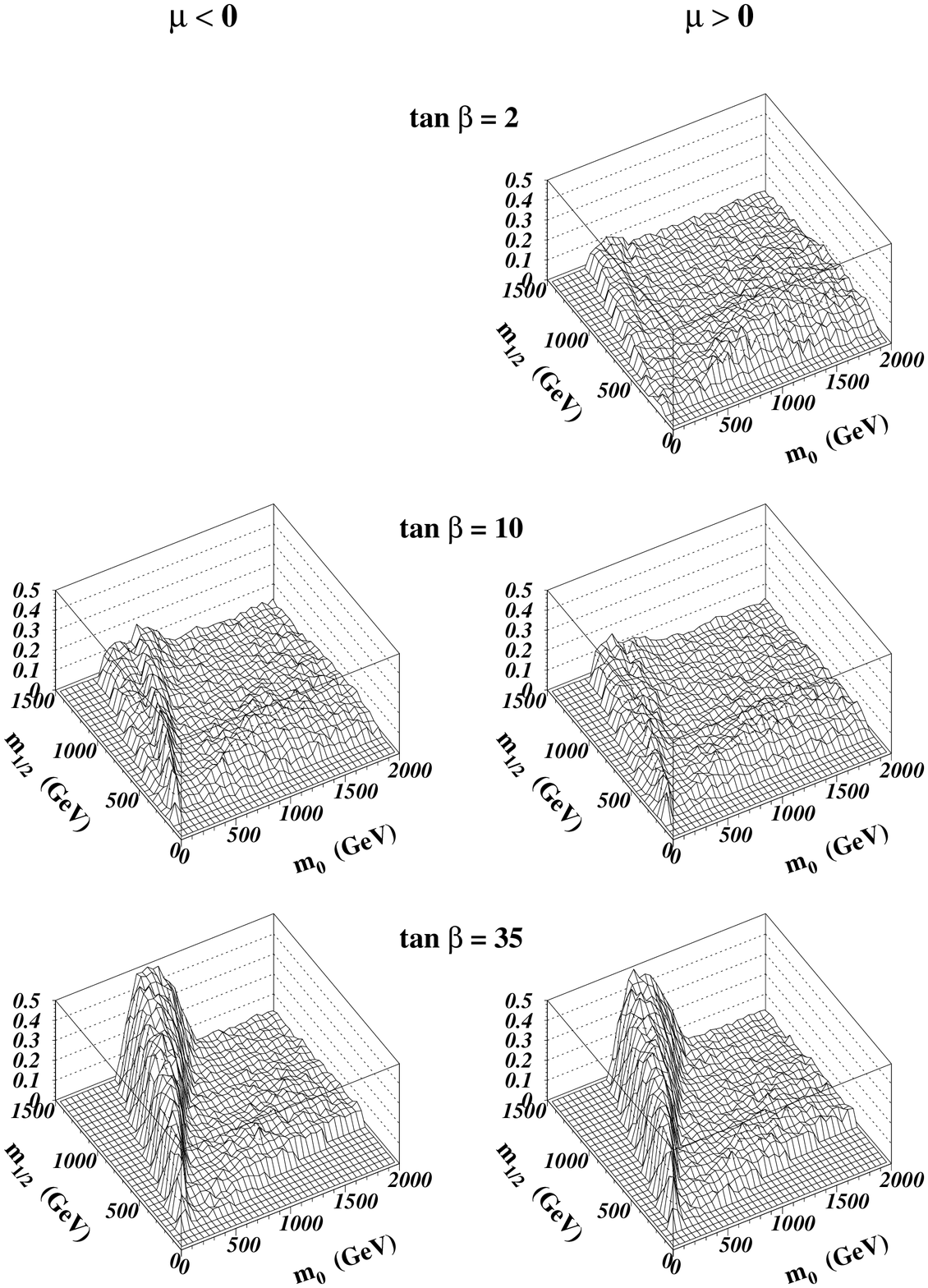}}
 \caption{ Same as in Fig.11, but for at least 2 taus.}
\end{center}
\end{figure}
\begin{figure}[hbtp]
\begin{center}
\vspace*{0mm}
\hspace*{0mm}\resizebox{0.99\textwidth}{22cm}
                        {\includegraphics{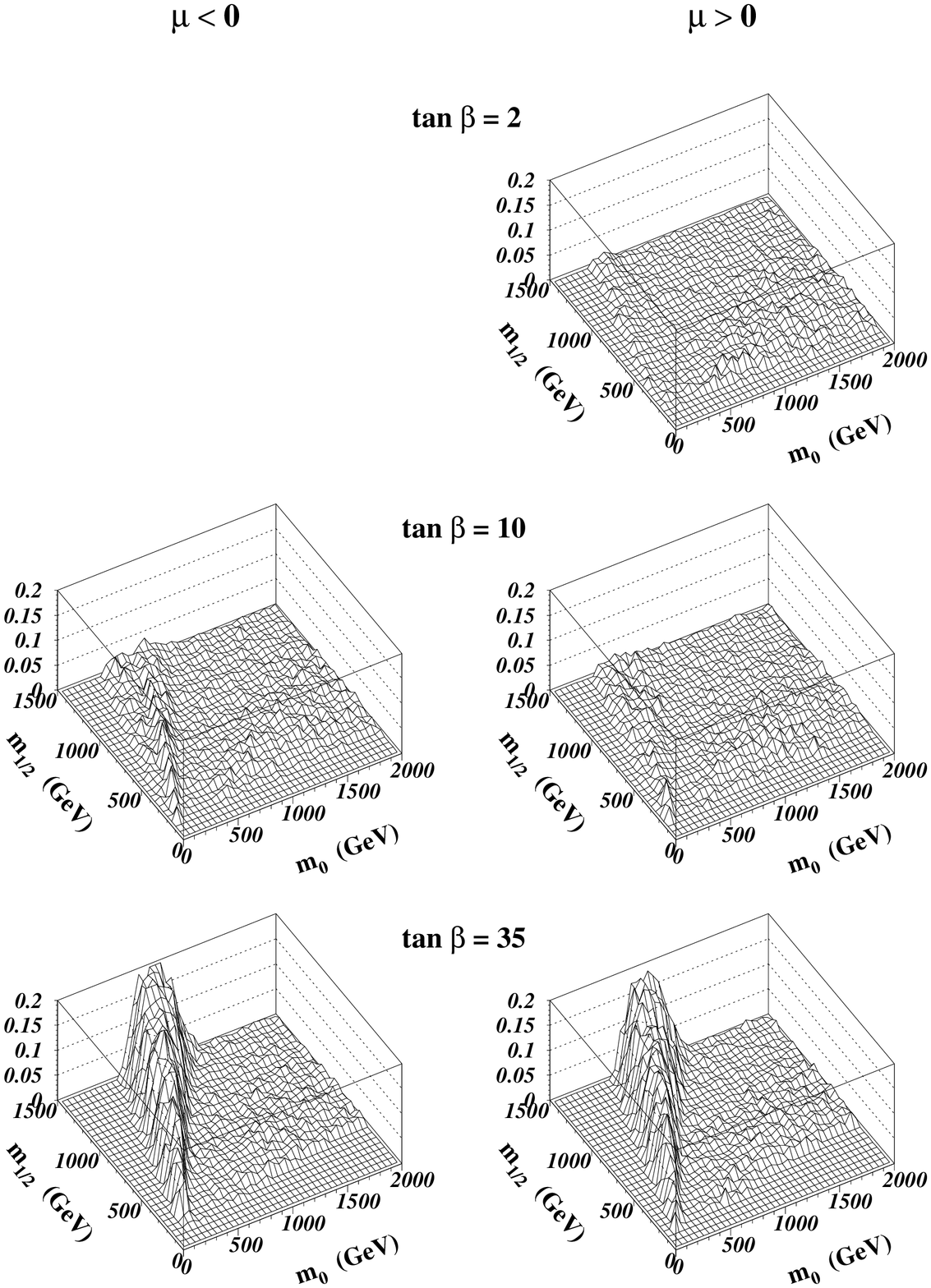}}
 \caption{ Same as in Fig.11, but for at least 3 taus.}
\end{center}
\end{figure}

\end{document}